\PassOptionsToPackage{hyphens}{url}

\documentclass[sigconf]{acmart}

\setcopyright{none} 
\acmConference[Anonymous submission to ACM MobiCom 2023]{ACM MobiCom}{TBD}{ TBD}
\acmYear{2022}

\acmBadgeR[]{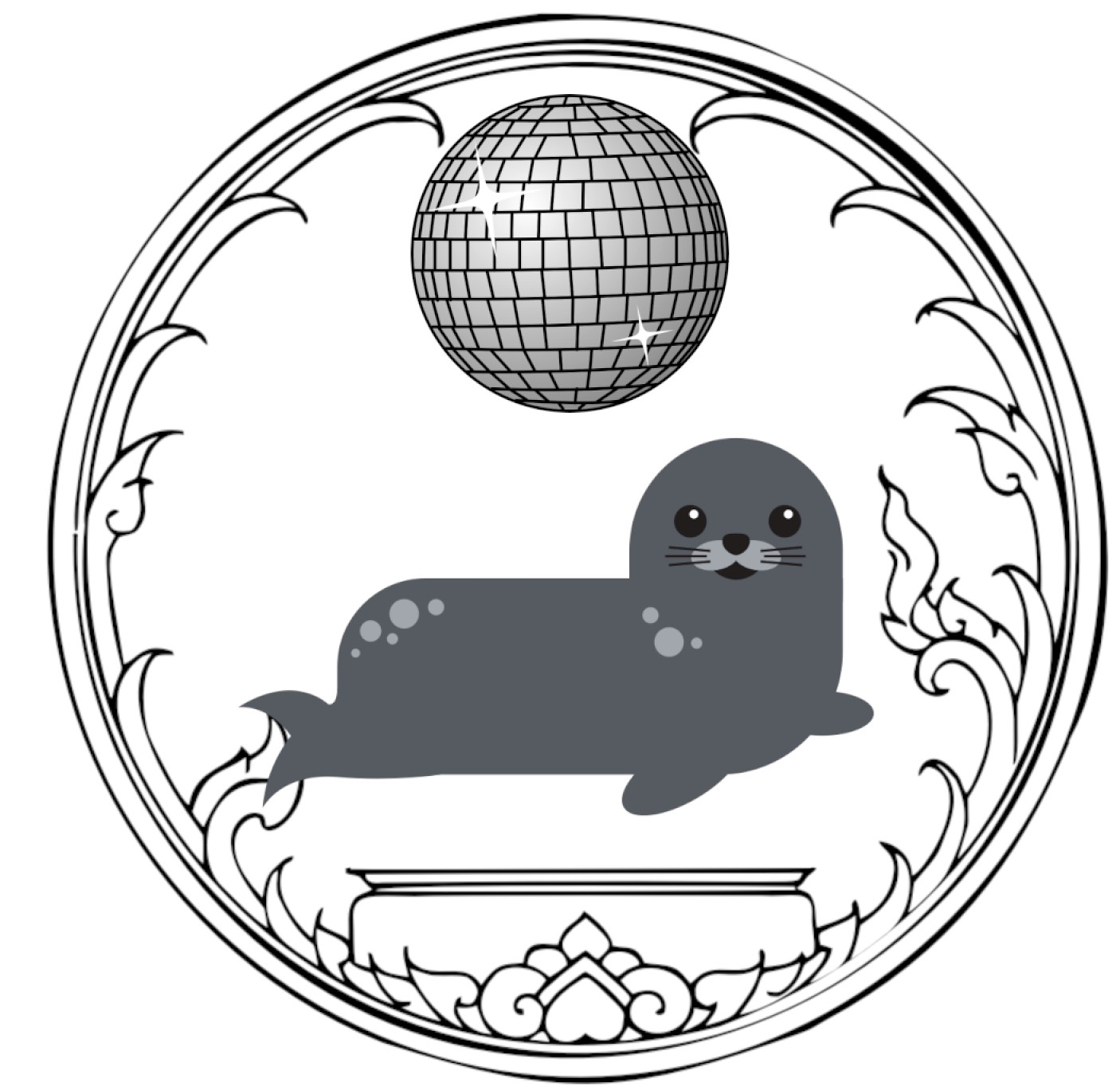}

\usepackage{epsfig,endnotes}
\usepackage{epsfig,endnotes}
\usepackage{xcolor}
\usepackage{eufrak}
\usepackage{amsthm}
\usepackage{amsmath}

\usepackage{amsfonts}
\usepackage{subcaption}
\usepackage{float}
\usepackage{algorithm}
\usepackage{algpseudocode}

\usepackage{booktabs}
\usepackage{multirow}

\newcommand\approach{SealClub}

\begin{document}

\date{}


\title{\approach: Computer-aided Paper Document Authentication }

\author{Martín Ochoa}
\affiliation{Department of Computer Science ETH Zurich}
\email{martin.ochoa@inf.ethz.ch}

\author{Jorge Toro-Pozo}
\affiliation{Department of Computer Science ETH Zurich}
\email{jorge.toro@inf.ethz.ch}

\author{David Basin}
\affiliation{Department of Computer Science ETH Zurich}
\email{basin@inf.ethz.ch}

\maketitle


\subsection*{Abstract}
Digital authentication is a mature field, offering a range of solutions with rigorous mathematical guarantees. Nevertheless, paper documents, where cryptographic techniques are not directly applicable,  are still widely utilized due to usability and legal reasons.

We propose a novel approach to authenticating paper documents using smartphones by taking short videos of them. Our solution combines cryptographic and image comparison techniques to detect and highlight subtle semantic-changing attacks on rich documents, containing text and graphics, that could go unnoticed by humans. We rigorously analyze our approach, proving that it is secure against strong adversaries capable of compromising different system components. We also measure its accuracy empirically on a set of 128 videos of paper documents, half containing subtle forgeries. Our algorithm finds all forgeries accurately (no false alarms) after analyzing 5.13 frames on average (corresponding to 1.28 seconds of video). Highlighted regions are large enough to be visible to users, but small enough to precisely locate forgeries. 
Thus, our approach provides a promising way for users to authenticate paper documents using conventional smartphones under realistic conditions.


\section{Introduction}

\paragraph*{Problem Context}

Forgery of printed documents is a widespread problem~\cite{financialforgery}. For example, it is believed that a single degree mill company sold over 200,000 fake educational diplomas worldwide for \$51 million in 2015 alone~\cite{diplomafake18}. Germany's Federal Criminal Police Office has detected about 70,000 document forgeries per year over the last decade~\cite{bka}. In 2018, a US prosecutor was found guilty of forging a judge's authorization for a wiretap~\cite{courtorders2}. As a consequence US legislation  has been introduced to combat counterfeit court orders~\cite{courtorders} by using digital signatures conforming to standards defined by the National Institute of Standards and Technology~\cite{NISTdigital}.  

Despite advances in digitalization, and mature solutions for digital signatures, printed paper is still widely used for documents of various degrees of sensitivity. In the US alone, 129 billion physical mails were sent in 2020~\cite{usps2020}. In enterprise environments, departments like human resources, legal, and accounting are still predominantly paper-centric ~\cite{whypaperenterprises}. Moreover, research on the impact of media on reading outcomes shows advantages for reading comprehension and efficiency when reading from paper as compared to reading from screens \cite{DELGADO201823,readingpaper1,readingpaper2}. It thus seems unlikely that paper will be completely replaced by electronic communication.

Traditionally, physical measures such as security paper, holograms, and watermarks, have been deployed to raise the bar against forging paper documents. Nevertheless,  forgery's high potential pay off motivates sophisticated and successful attacks~\cite{europolforgery}.

This situation raises a natural question: can we leverage standard security mechanisms from the digital world, such as cryptographic signatures, to prevent attacks on paper documents in the physical world? This question has inspired various proposals~\cite{eldefrawy2012hardcopy,li2015authpaper,wang2015cryptopaper, qrwithocr} such as  2D annotations (for instance QR codes) containing additional information, like signed hashes of the document's content. This annotation allows interested parties to verify a printed document's authenticity, assuming they possess the signer's public key. It is now also common to see documents with QR codes or URLs pointing to an authentic version of the document on the issuing institution's website~\cite{freiburgdegree,mumbaidegree,brazilapostille}. These approaches and other related work are discussed in more detail in Section~\ref{sec:relatedwork}.

Such techniques help preventing forgeries, however they have strong limitations. In most of these proposals, a human verifying a document's authenticity must still manually compare a printed document against an online version or against text stored in a QR code and displayed on a PC or smartphone. While this can help humans spot obvious forgeries, like those that drastically change a document's form or contents,  more subtle attacks where only a few characters are altered are challenging to spot manually. It is well known, for instance, that a single comma can alter the semantics of a sentence~\cite{10examples}. Consider, for example, the sentence ``Stop clubbing baby seals" versus ``Stop clubbing, baby seals". \footnote{Also a meme of internet fame.}

Alternatively, approaches have been proposed for automatic or semi-automatic comparison, such as~\cite{qrwithocr,andreeva2020comparison}. These techniques are tailored to pure text and use either OCR or machine learning. These are strong limitations since semantically meaningful attacks may modify charts, signatures, or other non-textual parts of a document. In addition, machine learning techniques are prone to adversarial attacks~\cite{attacksocr}. This creates additional attack vectors for which no robust generic countermeasures are currently known~\cite{chakraborty2021survey}. 

\paragraph*{Approach}
We propose a system to digitally authenticate printed documents. We call our approach \emph{\approach{}} and there are two main components to our solution: (i) cryptographic mechanisms are used to guarantee the authenticity and privacy of reference  document pictures and (ii) an image comparison algorithm is used to detect subtle forgeries on a short user-taken video of a paper document with respect to an authentic reference document. We use a video instead of a single picture for both usability and accuracy reasons: superimposing the outcome of the analysis on each video frame (augmented reality) provides users with immediate feedback and comparing multiple images against the reference image improves accuracy  and helps  filter out false forgery alarms.  

\begin{figure*}[h!]
    \centering
    \includegraphics[width=0.53\textwidth]{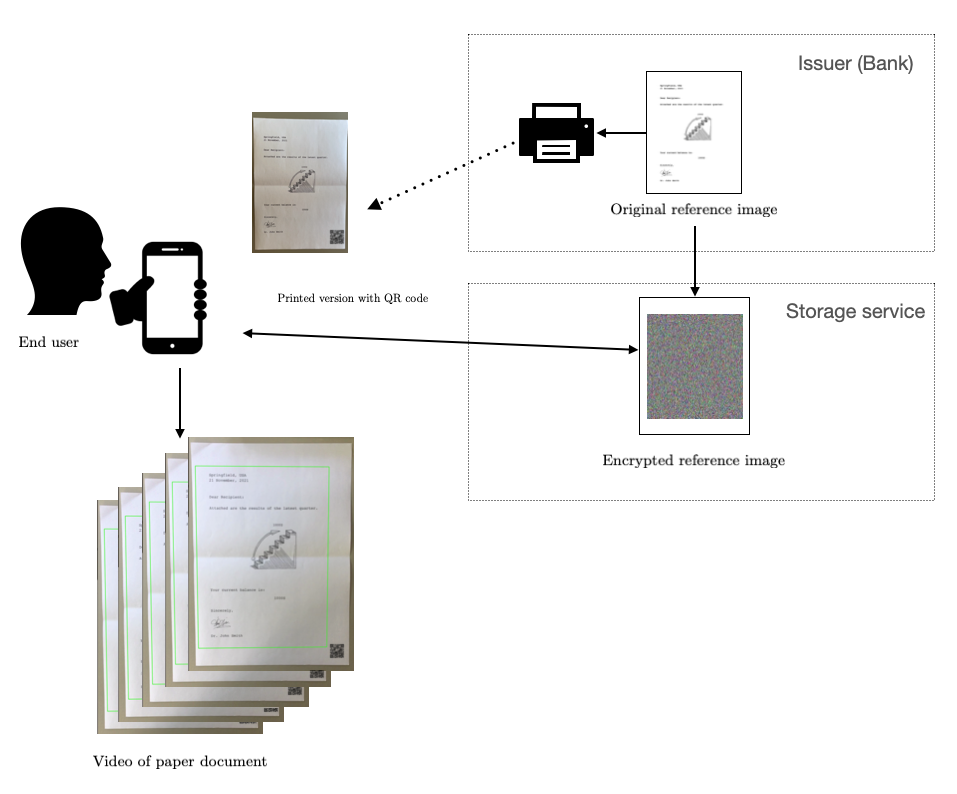}
    \caption{High level overview of \approach{}.}
    \label{fig:system}
\end{figure*}

Our image comparison algorithm is based on approximate similarity that automatically highlights noticeable differences between two document images. Our algorithm marks regions that differ by more than a pre-configured threshold and displays the differences to end users in a meaningful way, enabling them to make informed decisions about a document's authenticity. Crucially, our algorithm works in realistic conditions using off-the-shelf smartphone cameras and is tolerant to routine distortions, for example arising from document folding or adverse lighting conditions, thereby reducing false positive rates while maintaining high forgery detection rates.

To illustrate how this works and how cryptography is used, consider a scenario where a bank issues paper documents to customers, like account statements, as illustrated in Fig.~\ref{fig:system}. The bank has a public/private key pair.  With the private key, the bank signs a URL containing a digital copy of the document to be authenticated together with a hash of the image's content. The printed version of the document contains a QR code with this signature, the URL and the document's hash. Bank users receiving printed documents install an authentic bank app on their phone, which has pinned the bank's public key. With this key, the app can verify the signature in the QR and download the digital copy and check its authenticity. The app runs our image comparison algorithm, which compares the authentic digital copy of the reference against a short video of the physical document. As a result, the app either declares the physical document as authentic or highlights those regions that have been potentially tampered with. 


We have designed \approach{} to be secure even under strong attackers. As described in Section \ref{sec:system-description}, it uses multiple keypairs and other security measures to be secure against attackers who can compromise individual components or their private keys. Our approach provides message origin authentication and, as we will see in Section \ref{sec:security-analysis}, timestamp-based non-repudiation of origin under key revocation. It is relatively easy to extend our solution to offer additional features. For example, we may count, on the storage server, how many times a document has been authenticated. We do not, however, explore such extensions in this paper.

We also provide privacy guarantees by employing symmetric cryptography to protect against data breaches on the cloud storage that could expose potentially sensitive documents. Namely, \approach{} stores digital copies of the authentic documents encrypted with symmetric keys that are also included in the printed QR, but not stored by the issuer. These keys  can be used by end users to decrypt the retrieved documents from the remote server.

We formally model the protocol underlying \approach{} and  prove that it ensures document authenticity with respect to strong adversaries. We also implement our approach and evaluate its accuracy on videos of different document types, with and without forgeries, and taken with multiple devices. Our evaluation shows that our approach detects all forgeries with no false alarms. Moreover, for forgeries, the  highlighted regions are large enough to be easily seen by human users, but small enough to precisely pinpoint where the forgeries occur. 

\paragraph*{Contributions}

We are the first to provide a solution combining cryptography and image comparison to authenticate printed documents in the physical world. Moreover, our solution is secure against strong attackers and is  privacy-preserving, even in the presence of compromised components and server-side data breaches. We formalize our system design and provide formal guarantees on the cryptographic protocol that authenticates reference documents.  Finally, we implement our solution and show that it provides high assurance in real-world scenarios by an evaluation on 128 document videos. In our evaluation, all forgeries were detected with no false alarms after analyzing just 5.13 frames on average, corresponding to 1.28 seconds of video at a sampling rate of 4 frames per second.

\paragraph*{Organization}

The rest of the paper is organized as follows. We describe the problem statement, general requirements, and system and attacker models in Section~\ref{sec:requirements}. We present the design of \approach{} in Section~\ref{sec:system-description} and the associated security guarantees in Section~\ref{sec:security-analysis}. 
We evaluate and discuss limitations of our approach in Section~\ref{sec:evaluation}, review  related work in Section~\ref{sec:relatedwork}, and draw conclusions in Section~\ref{sec:conclusions}.

\section{Preliminaries}
\label{sec:requirements}
Before describing our technical solution in detail, we elaborate on the problem statement, the system and attacker models, and the technical requirements for solutions to the stated problem in this context. We also introduce relevant notation and give a first (ideal) definition of forgery and forgery detection.

\paragraph*{Problem Statement} \emph{Is it possible to build a  digital system that supports users in authenticating paper documents and recognizing forgeries?}

\subsection{Requirements}

\paragraph*{Functional requirements} Document issuers can share paper documents with users. Documents may be sent by postal mail, which may entail folding them in halves or thirds to fit in envelopes. Users can access a digital copy of the physical document they receive and automatically compare the physical and digital copies to detect differences using their standard, off-the-shelf smartphone. Frames in videos of documents taken using smartphones in uncontrolled scenarios may introduce lighting, focus, and geometrical distortions~\cite{distorsiondocuments}. For example, unfolded documents may fail to lie flat, which changes the geometry of the original (2D flat) digital document. Moreover the overall quality of pictures may change across smartphone models and users.

\paragraph*{Security and Privacy Requirements}  If an attacker modifies or replaces an existing paper document issued by a legitimate issuer before it reaches an end user, the phone-based app should alert the user to modifications in the forged document. Moreover, unauthorized parties should not be able to create new paper documents on behalf of a legitimate document issuer. If they attempt to do so, the app must inform end users of such spoofed documents. Only users in possession of a  legitimate paper document can retrieve the corresponding original digital counterpart. In case of a breach on the storage service, which stores the original documents in digital form, attackers should not be able to learn the content of the documents unless they possess the corresponding paper documents. In the event of an issuer's key compromise, users in possession of issued documents before the key is revoked should still be able to authenticate them (non-repudiation of origin under key revocation). Moreover compromise of individual system components (issuing or storage) should not suffice to successfully create forgeries.

The system must allow for key revocation in case of compromise of private keys. In this case, user must be able to authenticate documents issued prior to revocation (non-repudiation of origin under revocation \cite{nonrepudiation2}).

\paragraph*{Usability Requirements}
The system must tolerate a wide range of uncontrolled distortions on the video taken by end users with their smartphones. This means that the system must have a low number of false alarms on legitimate documents, but still provide high accuracy in detecting forgery. This is challenging because, as discussed in the introduction, a semantic-changing forgery could be as (visually) small as a single comma, added or removed.

\subsection{System and Attacker Model}
\label{sec:sys-n-attacker}


\paragraph*{System Model} There are two kinds  of legitimate system users: \emph{issuers} and \emph{end users}. Issuers have an insecure channel, such as postal mail, for paper documents but can authentically distribute their public keys to  users. 
Users connect to the storage service to securely retrieve encrypted copies of a document's digital version.  Documents have a secure timestamp that is provided by a time-stamping service, and which allows end users to check when a document was created.

Users also visually compare a short video of the paper document and the (original) digital version retrieved from the issuer's servers.  
 This comparison is assisted by an app that automates the visual highlighting of (potential) differences between frames in the video and the reference image of the document. 

\paragraph*{Attacker Model} Attackers may attempt to modify existing, legitimate, paper documents, by hijacking them in transit to end users or by creating new documents on behalf of issuers. To do so, they may use any forgery means, digital or physical, and may perform subtle attacks to increase their success chances. However, crucially, a forgery must be visible and semantic changing. 

Note that it is hard to define \emph{semantic changing} precisely in general, since this will vary from document to document. For instance, for some documents, like those referring to our baby seals, small punctuation changes suffice to change the document's intended meaning. For other documents, changing the signature or a portion of a graph representing sales results may be a semantic changing attack. However, changing a pixel or a region that is smaller than a punctuation mark may not change the document's intended content for human readers, but will simply appear to be noise.

Attackers may compromise individual components such as the issuer or the storage service (see Fig. \ref{fig:system}) and obtain their private keys. However, we assume end users can download an authentic app on their smartphones and can update it securely; thus we rule out fake or compromised app attacks.


\subsection{Video processing}

In order to improve accuracy and to give users real-time feedback on the quality of the pictures the algorithm receives, our solution works with short videos of paper documents rather than single pictures. However, the comparison with respect to the reference document will be performed on the individual frames of a video. 

Given a video and a frame-by-frame comparison, we can define a notion of \emph{analysis convergence}. Namely, for a fix number $k$, we compare the result of the analysis on the last $k$ frames of the video. If the results are identical (frames are authentic or they contain forgeries in the same positions) then we stop the analysis and report the result. For most of Section \ref{sec:computervision}, we will thus describe single image (frame) comparison against the reference document.

\subsection{Strict Forgery Detection}

For the rest of the paper, we will consider grayscale documents and begin with some notation.  Let $n, m\in \mathbb{N}$ and $\mathcal{D} =[0,1]^n \times [0,1]^m$. A digital document $d \in \mathcal{D}$ is an $n\times m $ grayscale matrix, where $[0,1]$ is a fixed-point representation of fractions between $0$ and $1$. Here we use the computer vision convention for pixel intensity where $0$ stands for black, $1$ for white, and in-between values are a gray scale. In the following, it will be useful to consider rectangles within an image, which we denote by two coordinates 
$(x_1, y_1),(x_2,y_2) \in [0,n]\times [0,m]$ such that $x_1 < x_2$ and $y_1 < y_2$.

A (perfect) printing function $p: \mathcal{D} \to \mathcal{P}$ transforms the digital document into a paper version $\mathfrak{p}$ that faithfully preserves proportion and intensity. Ideally, there exists an inverse function $p^{-1}: \mathcal{P} \to \mathcal{D}$ such that $d = p^{-1}(p(d))$. For instance, a perfect printing of $d$ that is perfectly scanned should be pixelwise identical  to the original $d$.

Although forgery can happen at the digital or physical level, we can abstract away from where it occurs by comparing two digital images $d$ and $\hat{d}$, where $\hat{d}$ may be $p^{-1}(p(\hat{d}))$ (a scan of a document that was possibly digitally modified and then scanned) or $p^{-1}(\widehat{p(d)})$ (a scan of authentic printed document that was possibly physically modified).

\begin{definition}[Forgery] \label{def:forgery}
Let $d,d' \in \mathcal{D}$. If $d$ is an authentic image, we say \emph{$d'$ is forged} if there exists at least one position $(i,j) \in [1,n] \times [1,m]$ such that $d(i,j) \neq d'(i,j)$. 
\end{definition}

A \emph{forgery locator} is an algorithm that computes the regions that have been modified. A neighbour of a position $(x,y) \in \mathbb{N}^2$ is a position $(x',y') \neq (x,y)$ such that $max(|x-x'|,|y-y'|) \leq 1$.   In a \emph{connected set} $C \subseteq \mathbb{N}^2$, all positions $(x,y) \in C$ have at least one neighbour in $C$.

\begin{definition}[Ideal forgery locator]\label{def:forgery_detector}
Let $d, d'\in \mathcal{D}$ be an authentic image and a potential forgery of $d$, respectively. Then an \emph{ideal forgery locator} $F(d,d')$ returns a list of disjoint detected differences $r_1, \dots, r_k$, such that each $r_i$ is the smallest rectangle containing a connected set of positions $x_j,y_j $ with $d(x_j,y_j) \neq d'(x_j,y_j)$ and the  empty set otherwise.
\end{definition}

These definitions correspond to the strictest security guarantees in an ideal scenario where even a difference in one single pixel constitutes a forgery. In practice, however, it will be impossible to enforce such strict guarantees when working with printed paper since smartphone-scanned images of printed documents suffer from distortions, as discussed in Section~\ref{sec:requirements}. Hence, we must relax the definition of forgery and a forgery locator to tolerate some differences as we explain in the following section.

\section{\approach{} Design}
\label{sec:system-description}

\approach{} combines algorithms for image comparison with  cryptographic measures to support end users in determining a document's authenticity.  We describe this in detail  in the next sections.

\subsection{Image Comparison Algorithm}
\label{sec:computervision}

Next describe our image comparison algorithm which is at the core of our approach. 
We first motivate the kind of algorithm required, based on the requirements of Section~\ref{sec:requirements}, before describing how it works.


\subsubsection{Approximate Similarity} 

In practice when dealing with physically printed documents there will inevitably be pixel-wise differences. 
Definitions~\ref{def:forgery} and~\ref{def:forgery_detector}, although appealing from a security point of view, are too strict in practice as they will result in forgeries being detected even on pictures of documents that have not been modified (false alarms). 



\begin{figure*}
    \centering
    \includegraphics[width=0.75\textwidth]{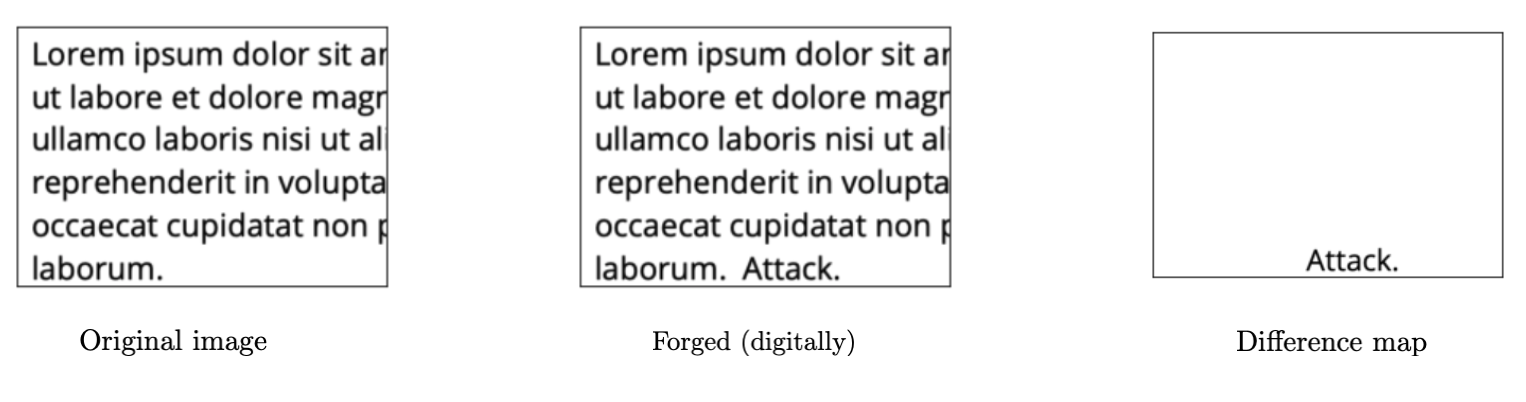}
    \caption{Difference map between authentic and forged (word addition). }
    \label{fig:ex1diff}
    
\end{figure*}

In order to better capture meaningful differences between images, we introduce the notion of a \emph{difference map}.

\begin{definition}[Difference map] The \emph{difference map} of two images $d$ and $d'$ is the matrix, also in $\mathcal{D}$, defined by:
$$\text{diff}(i,j) =|d(i,j) - d'(i,j)|,$$
\noindent for all positions $(i,j) \in [1,n] \times [1,m]$.
\end{definition}

Intuitively, the difference map is a matrix containing the visual differences between two images, with more glaring differences having higher intensity (i.e.~a black pixel against a white one). For instance, Figure~\ref{fig:ex1diff} shows how adding a word (digitally) to a textual image can be seen in the difference map.

\begin{figure*}
    \centering
    \includegraphics[width=\textwidth]{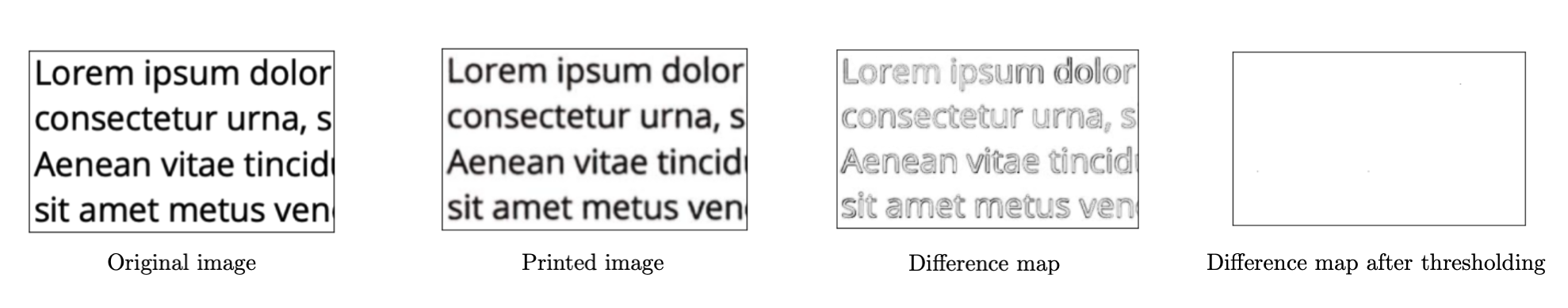}
    \caption{Differences between an authentic digital picture and its printed version. The last image is the result of applying a threshold $\delta = 0.15$ to the difference map. }
    \label{fig:ex2diff}
    
\end{figure*}

When comparing a high quality picture of a printed document (paper lies completely flat, good lighting and focus, image preprocessing) against the original, we still get some small visible differences in the difference map. These are, however, easy to filter using intensity thresholding, as can be seen in Figure~\ref{fig:ex2diff}.

To account for a tolerance to small differences, we relax our definitions of forgery and a forgery locator.

\begin{definition}[$(\delta,\tau)$-detectable forgery]\label{def:tdforgery}
Let $d,d' \in \mathcal{D}$. A $(\delta,\tau)$-\emph{detectable forgery} is a connected set of positions $pos_1,\dots,pos_k \in [1,n] \times [1,m]$ in the local difference map $M$ of $d$ and $d'$   such that $M(pos_j) < \delta$, for each $1\leq j \leq k$ and $k >\tau$ . 
\end{definition}

The motivation behind the choice of parameters $\delta$ and $\tau$ is that differences with low intensity ($\delta$) in the difference map correspond to less visible differences, such as blurred borders of text, and small regions ($\tau$) are usually noise. The concrete values of $\delta$ and $\tau$ must be set in a concrete implementation and may vary depending on a document's intrinsic feature, for instance the font size or character set. These values are defined by document issuers and are embedded in the paper document's QR code as we will discuss in the next section. Note that setting $\delta = 1$ and $\tau = 1$ is equivalent to the stricter Definition~\ref{def:forgery_detector}.

A $(\delta,\tau)$-forgery locator points to specific regions that have been modified.

\begin{definition}[$(\delta,\tau)$-forgery locator]
\label{def:tdforgery_detector}
Let $d, d'\in \mathcal{D}$. Then a $(\delta,\tau)$-\emph{forgery locator} $F(d,d')$ returns a possibly empty list of detected differences $r_1, \dots, r_k$, where $r_i$ is the smallest rectangle containing a $(\delta,\tau)$-forgery.
\end{definition}

\begin{figure*}[tb]
    \centering
    \includegraphics[width=\textwidth]{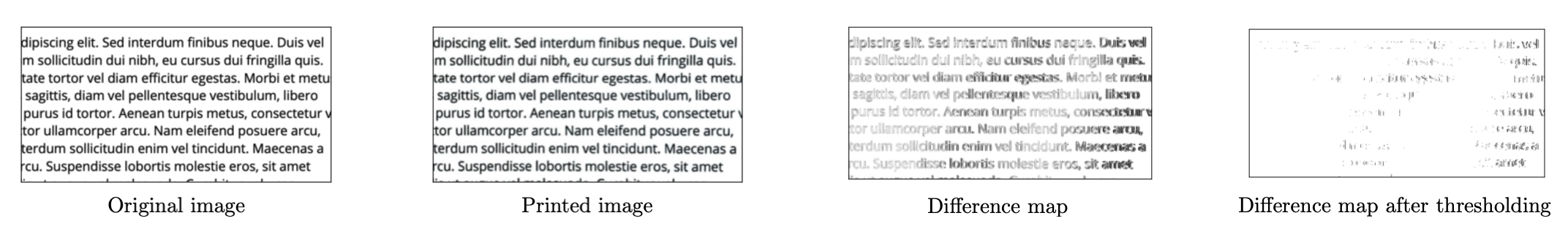}
    \caption{Differences between an authentic digital picture and its printed version when the printed version is slightly folded. The last image is the result of applying a threshold $\delta = 0.15$ to the difference map. }
    \label{fig:ex3diff}
    
\end{figure*}

In practice $d$ will be an image of the authentic document and $d'$ a picture of the paper document to be authenticated. We emphasize that we seek a practical solution where $d'$ can be produced by a standard smartphone under normal conditions. Solutions requiring the careful use of high-quality flatbed scanners on pristine paper documents would, for example, be inappropriate for general, widespread use. A practical challenge then is that uncontrolled smartphone pictures of a printed document have unpredictable distortions as in Figure~\ref{fig:ex3diff}. 

Although Definition~\ref{def:tdforgery_detector} is more tolerant to slight differences between images, while still providing some security guarantees, examples such as Figure~\ref{fig:ex3diff} are  challenging to handle, given that the differences are large enough to be considered potential attacks. In the next section, we discuss how we tackle these challenges while still providing clearly defined security guarantees.

\subsubsection{An Iterative Forgery Detection Algorithm} 

An important precondition for document comparison is that the two pictures $d$ and $d'$ are aligned. Alignment is non-trivial when using smartphone pictures as they  are likely to have angle and perspective differences with respect to the authentic image.

\begin{figure*}[t]
\centering
\begin{subfigure}{.35\textwidth}
  \centering
  \includegraphics[width=\linewidth]{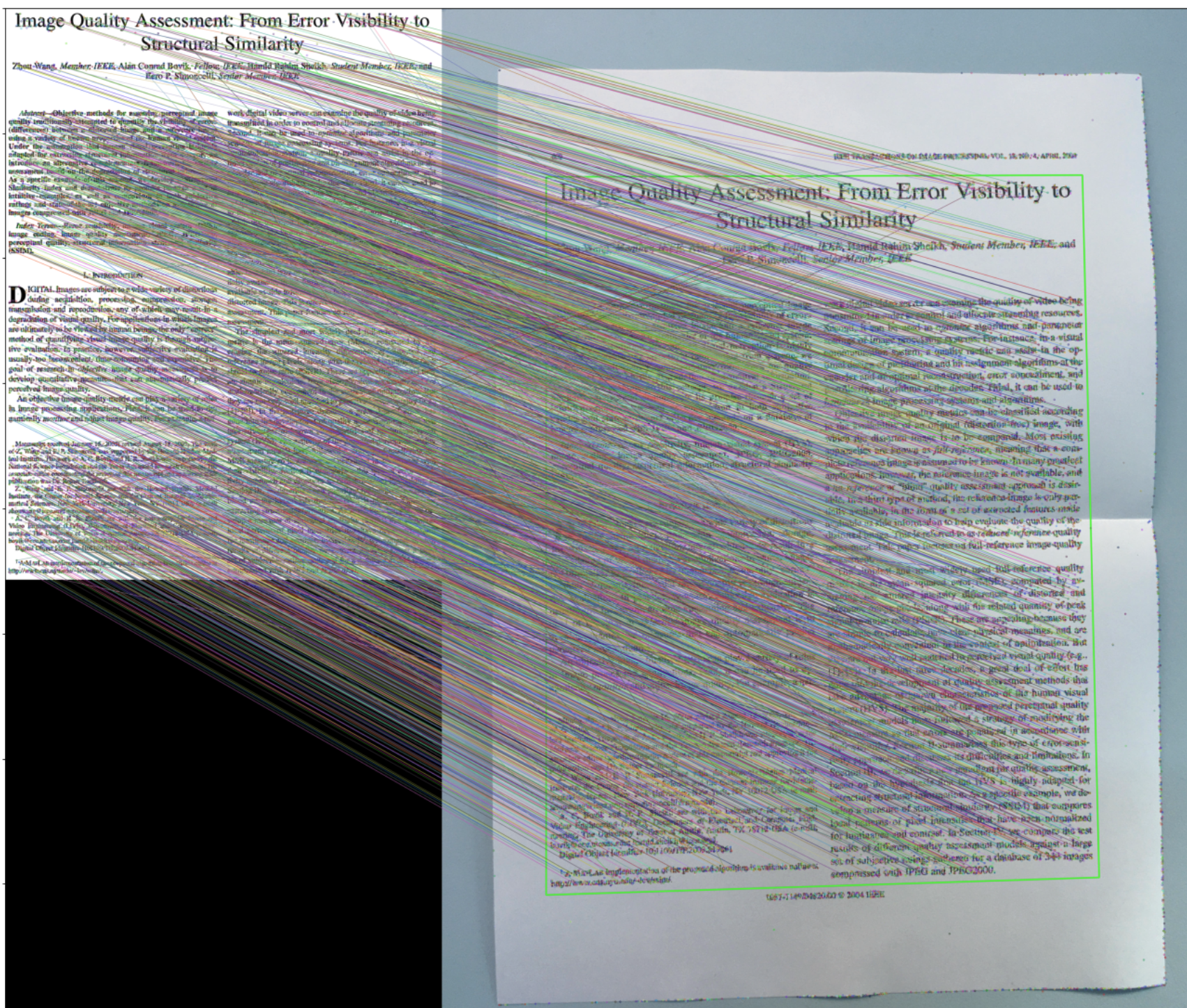}
  \caption{Descriptive feature matching and estimation of perspective}
  \label{fig:matching}
\end{subfigure}%
\begin{subfigure}{.45\textwidth}
  \centering
  \includegraphics[width=.5\linewidth]{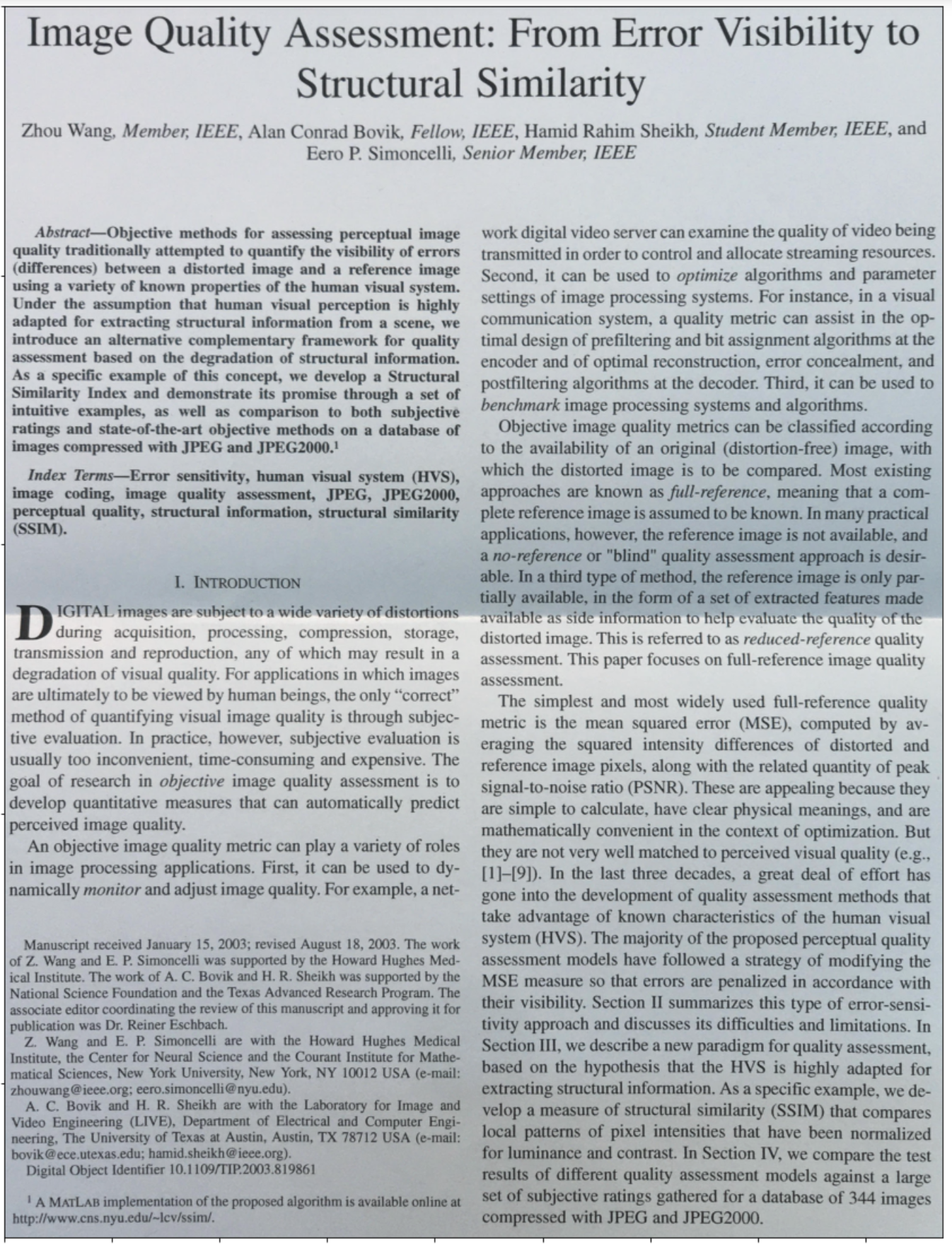}
  \caption{Result of homography computation
  }
  \label{fig:result}
\end{subfigure}
\caption{Result of object matching and perspective adjustment using computer vision techniques.}
\label{fig:homography}
\end{figure*}

A commonly used technique to rectify such differences is to apply a \emph{homography} computed over the document's estimated corners~\cite{geetha2013automatic}. This technique is widely implemented in smartphone applications for document scanning such as Google Lens and Microsoft Lens. In our setting, we have an additional advantage when computing an accurate homography since we have access to the  original image. We can thus use computer vision techniques to recognize known objects in new scenes. Popular algorithms to find descriptive points of an object are, for instance, scale invariant feature transform (SIFT)~\cite{lowe2004sift} and Oriented FAST and Rotated BRIEF (ORB)~\cite{orb}. We can then search for matches of such points in two images using automatic feature matching~\cite{muja2009fast}. This allows us to better estimate the location of the authentic document in the printed document scan and then apply a homography to align both pictures. Figure~\ref{fig:homography} illustrates the combination of both techniques, which is an instance of general object finding~\cite{objectmatching}.

Our approach is as follows. Folds in the printed document may induce distorsions that cannot be corrected with a global homography technique, given that they occur in 3D space as illustrated in Fig.~\ref{fig:ex3diff}. However, if we zoom in to the neighborhood of a detected potential difference, we can attempt to recompute a homography in this limited region and check for differences. The rationale for this is that locally, in the neighborhood of a potential difference, the perspective is more consistent and thus more likely to be corrected.

\begin{figure*}[tb]
    \centering
    \includegraphics[width=0.9\textwidth]{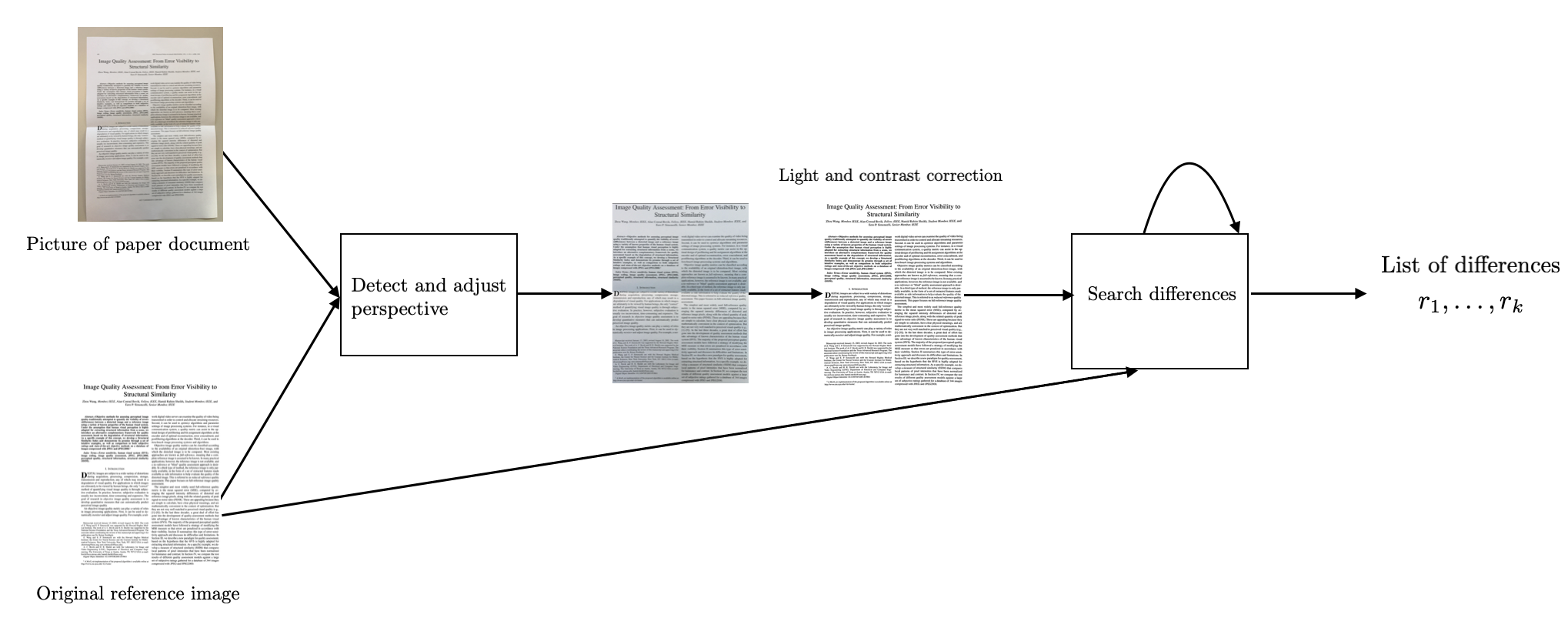}
    \caption{High-level view of proposed iterative approach to build a forgery locator.}
    \label{fig:algorithm}
    
\end{figure*}

\begin{algorithm}[h]
\caption{Iterative Forgery Detection Algorithm}\label{alg:algo}
\raggedright
\textbf{Input:} Images $d,d' \in \mathcal{D}$ to be compared\\
\textbf{Output:} List of differences $\Delta$\\
\begin{algorithmic}[1]
\State $d' \gets \mathtt{find}(d,d')$
\If{$d'$ is Null}
    \State $\Delta \gets \{ (0,0),(n,m) \}$
   \State \Return $\Delta$ \Comment{Documents are very different}
\EndIf
\State $d' \gets \mathtt{preprocess}(d')$ \Comment{Lighting and contrast}
\State $\Delta \gets \emptyset$
\State $\Delta' \gets \mathtt{get\_differences}(d,d')$
\While{$\Delta \neq \Delta'$}
    \State $\Delta \gets \Delta'$
    \State $\Delta' \gets \emptyset$
 \For{$r \in \Delta$} 
 \State $\sigma \gets \mathtt{neighborhood}(r)$
 \State $z \gets \mathtt{find}(d(\sigma),d'(\sigma))$
 \If{$z$ is Null}
    \State $z \gets d'(\sigma) $ \Comment{Region is different}
\EndIf    
 \State {$\Delta' \gets \Delta' \cup \mathtt{get\_differences}(d(\sigma),z) $} 
 \EndFor
 
\EndWhile
\State \Return $\Delta'$
\end{algorithmic}
\end{algorithm}

We present our forgery detection algorithm in Fig.~\ref{fig:algorithm}. First we compute descriptive points and look for a homography, controlling a rotation up to the angle $\phi$. We then perform typical preprocessing steps similar to document scanning to improve light and contrast. Afterwards we perform a first approximate similarity pass on the document obtaining a list $\Delta$ of potential differences. We iterate over each difference and compare again a neighborhood $\sigma$ of the difference in the authentic image against a neighborhood of the image to verify. The comparison involves recomputing the descriptive points and recomputing the homography.   We then recompute the approximate similarity difference map in the region $\sigma$ and store the newly found differences (if any). If no good homography is found, which can happen for instance if the printed document contains a region visually very different from the original (modified or new text/images), then we directly compare the two corresponding regions. The process can be repeated until a fixed point is reached. 

Pseudocode for our algorithm is given in Algorithm~\ref{alg:algo}. In this presentation, we assume that the parameters $\tau$ and $\delta$ are implicit to the function $\mathtt{get\_differences}$, that the size of neighborhood is implicit to $\mathtt{neighborhood}$, and that $\phi$ is implicit to $\mathtt{find}$.

\begin{definition}[Similarity]
When the output of Algorithm~\ref{alg:algo} is the empty list (no forgeries found), we say that  $d'$ and $d$ are $(\tau,\delta,\phi,\sigma)$-\emph{similar}.
\end{definition}

\paragraph{Video processing} The algorithm's output is thus a list of visual annotations containing potential differences. Some of those differences may be false alarms, which occur when picture distorsions cannot be automatically corrected. However, as discussed in Section~\ref{sec:requirements}, we used the output of single frames in the context of a video containing multiple frames to reach a verdict. 

\begin{definition}[Frame coherence] Let $k \in \mathbb{N}$. Let $\mathcal{V}$ an ordered list of documents (frames) $f_0,\dots,f_l$ for $l\geq k$. A set $F$ of $k$ consecutive frames  $f_i,\dots,f_{i+k}$ is \emph{$(\tau,\delta,\phi,\sigma)$-coherent} if the output of Algorithm~\ref{alg:algo} on each frame in $F$ is the same.
\end{definition}

Even if false alarms can arise due to the lighting or focus conditions in a given frame, they are less likely to occur in several consecutive frames. We will then converge to a given analysis after finding $k$ consecutive coherent frames. This turns out to be surprisingly effective as we show empirically in Section \ref{sec:evaluation}.

\subsection{Securing the Reference Document} 
\label{sec:digitaltrust}

Our forgery detection algorithm takes a reference image as input and it is thus essential to  ensure this image's authenticity. We refine the high-level architecture of Fig.~\ref{fig:system} to highlight the various components and the communication channels between them in Fig~\ref{fig:system2}. 
The key point of this construction is that documents printed by issuers are time-stamped and have a signed QR on them that includes both the URL and the hash of the original document's content; hence both are protected from modification and spoofing, as well of all relevant parameters. Documents signed with a revoked key are only authenticated by the app if the timestamp is older than the revocation date.

We start by describing the setup assumptions and afterward the component's behaviours. 

\subsubsection{Setup assumptions}

First, it is necessary to distribute keys such that the parties involved can issue and verify documents.  As usual, we assume an issuer $I$ has generated a pair of asymmetric keys $(k_I,K_I)$, called \emph{issuer keys}, and can sign digital documents using its private key $k_I$. Moreover, a time-stamping service has generated a key pair $(k_{TS}, K_{TS})$ of \emph{time-stamping keys} used to provide secure timestamps. 

The end user app must possess the  authentic public keys of the issuer and the time-stamping service. There are various standard ways how these public keys can be authentically distributed to end users. First, as discussed in the example in the introduction, users can download a  legitimate app from a trustworthy app store, and the app ships with the pinned public keys. Alternatively, public keys can be scanned, in the form of a QR, at a public physical location that is trusted by end users. For instance, if the issuer $I$ is a bank, then it can have a QR code exhibited in its offices for users to scan. %
Also, users can scan $I$'s public key in a QR code hosted on a secure website provided by $I$, as is sometimes done when enrolling for software-based two factor authentication. 

\begin{figure*}[ht]
    \centering
    \includegraphics[width=0.7\textwidth]{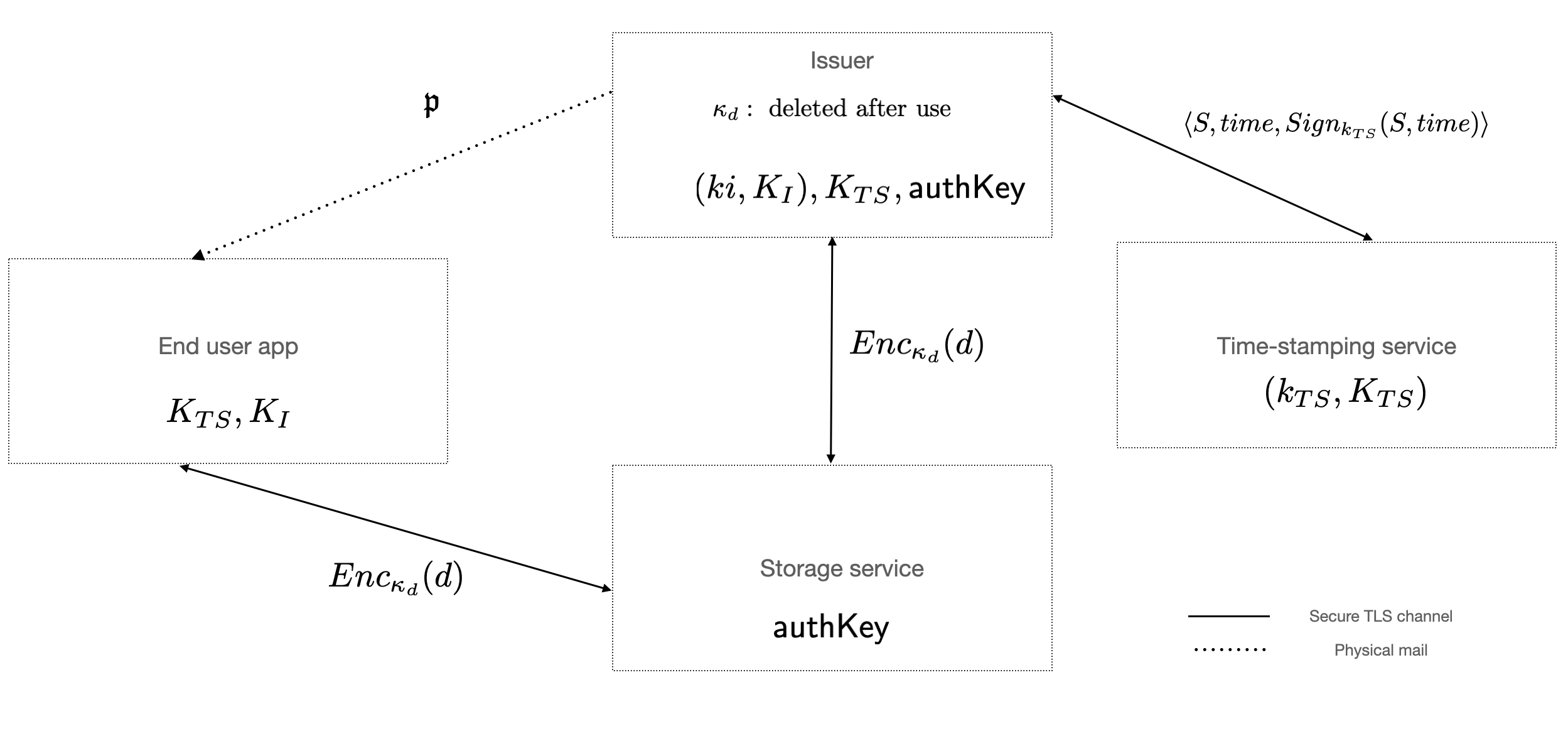}
    \caption{Component overview of \approach{}.}
    \label{fig:system2}
\end{figure*}

As indicated in Figure~\ref{fig:system2} we depict connections with the storage service and time-stamping service are to be secured using TLS. This also requires that these services have generated public-private key pairs and the app has access to the respective public keys. These public keys can also be pinned in the app, or the WebPKI can be used. It is necessary however that an issuer authenticates itself with the storage server to store documents and for this either a client-side certificate can be used, or the issuer can be authenticated over the TLS channel using a pre-distributed credential \textsf{authKey} for this purpose. For simplicity of exposition we assume the latter in our account below.


\subsubsection{Role descriptions}

We next describe the roles of the different system components. 

\paragraph{Issuer}

 To issue a new document $d$, issuer $I$ first generates a symmetric key $\kappa_d$ for encryption. Afterwards it creates the  QR code $q$ to be printed together with document $\mathfrak{p}=p(d)$. The QR can be placed either on the document's bottom or on its backside. After generating the QR code an adding it to $\mathfrak{p}$, the issuer deletes both $\kappa_d$ and $q$.

The QR code  contains the following data:
\begin{itemize}
    \item A URL $U$ pointing to the encrypted reference document $E=Enc_{\kappa_d}(d)$.
    \item A hash of the original document $h(d)$.
    \item The symmetric key $\kappa_d$ to decrypt $E$.
    \item (Optional) Fallback text $T$ describing the document for offline verification.
    \item The values of $\tau$ , $\delta$, $\phi$, and $\sigma$ for this document.
    \item A signature $S  = Sign_{k_I}(h(U,h(d),\kappa_d,T, \tau, \delta, \phi, \sigma))$ produced by the issuer.
    \item A secure timestamp $t$ of $S$
    \item The fingerprint of the issuer's public key.
\end{itemize}

Here, $U$ corresponds to the URL hosting $Enc_{\kappa_d}(d)$ in the storage service component, with which $I$ can authenticate using  \textsf{authKey} and build a secure channel using the storage service's TLS certificate. The secure timestamp is a 3-tuple $t = \langle S, time,Sign_{k_{TS}}(S,time)  \rangle$, consisting of the signature $S$, the current time, and a signature with $k_{TS}$ is obtained through a time-stamping service. $I$  shares the paper document $\mathfrak{p}$ together with the QR $q$ with the end user.

Table~\ref{tab:lengths} summarizes the storage requirements for the data encoded in the QR code, totaling 534 bytes (i.e. 1,068 hexadecimal digits: 0–9, A–Z). This amount of data fits in a QR code that can be easily scanned by ordinary smartphone cameras.

\paragraph{End user app} Upon receiving a document $\mathfrak{p}$ containing a QR code $q$, the application running on the user's smartphone looks up in its local storage the corresponding public key $K_I$ using the fingerprint, and then verifies the signature $S$. It also verifies the content of the timestamp $t$ using the public key  $K_{TS}$ of the time-stamping service. It then checks whether the time used in the timestamp and the issuer key used for signing the document $K_I$ are coherent with respect  to the Certificate Revocation List. That is, if the key was revoked, whether the timestamp is older than the revocation date. If so, the app then downloads the encrypted document $E$  establishing a secure channel with URL $U$ (using the pinned key $K_S$ of the storage service). It then decrypts  $E$ with the symmetric key $\kappa_d$ and verifies the hash of the original document. If the hash verifies, $E$'s plaintext is authentic and the app then  passes the plaintext to the forgery detection algorithm described in Section~\ref{sec:computervision}. Alternatively, if internet connectivity is not available, the app can display the text $T$ to the user and perform an Optical Character Recognition (OCR) on the document as a fallback mechanism.

\paragraph{Storage service} The storage service publicly hosts encrypted documents that can only be uploaded by the issuer. For this, issuers must register with the storage service setting up a shared credential \textsf{authKey} to authenticate subsequent uploads.

\paragraph{Time-stamping service} The time-stamping service receives a signature $S$ from the issuer to be used in the QR generation, and computes a secure timestamp triple $t = \langle S, time,Sign_{k_{TS}}(S,time)  \rangle$. Anyone possessing the public key $K_{TS}$ can verify that $S$ was produced at time $time$.

\subsubsection{Revocation}

An issuer's private key can be compromised, in which case documents subsequently signed with this key should no longer be considered authentic.  Revoked keys are stored, together with the time of revocation, in a publicly accessible Certificate Revocation List (CRL). This CRL is maintained by a trusted entity, which could be for example the issuer, the storage service provider, or a national certificate authority. The replacement of the revoked keys can be done on the app update. 

In our context, if an issuer's private key were revoked, it is desirable that documents that were issued pre-compromise should be trusted and verifiable in order to avoid the necessity of reissuing them. This property, called non-repudiation of origin, can be achieved using secure timestamps \cite{nonrepudiation1, nonrepudiation2}, which are provided by the time-stamping service. If the private key of the time-stamping service is also compromised, then non-repudiation cannot be achieved~\cite{nonrepudiation2}. However in Section \ref{sec:security-analysis} we will discuss how our system storage service provides a last line of defense against this threat.

\subsubsection{Document Privacy} 

In \approach{}, the original digital documents should be available online for end users to verify printed documents. Given that this poses the risk of a data breach, we  store those copies encrypted with a uniformly random symmetric key $\kappa_d$, unique for each document $d$. As discussed above, this key, generated by the document issuer, is stored in the printed document's QR code, and then deleted after printing. Only users in possession of the printed document or copies thereof can decrypt the original document stored in the storage cloud. 


\begin{table}[t]
    \centering
    \renewcommand{\arraystretch}{1.2}
    \begin{tabular}{c c c}
         \hline\hline
         \multirow{2}{*}{\bfseries Element} & \bfseries Max. length & \multirow{2}{*}{\bfseries Using}\\
         & \bfseries (bytes) &\\
         \hline
         $U$ & 100 & 100-char URL\\
         $h(d)$ & 32 & SHA 256\\
         $\kappa_d$ & 32 & 256-bit AES key\\
         $T$ & 200 & 200-char text\\
         $\tau$, $\delta$, and $\phi$ & $3 \times 2 = 6$ & 2-byte integers\\
         $\sigma$ & $2 \times 2 = 4$ & 2-byte integers\\
         $S$ & 64 & 256-bit ECDSA signature\\
         $K_I$'s fingerprint & 32 & SHA 256\\
         $time$ & 4 & 4 byte timestamp\\
         $S'$ & 64 & 256-bit ECDSA signature\\
         \hline
         \bfseries Total & \bfseries 538 \\
         \hline\hline
    \end{tabular}
    \caption{Estimated maximum lengths of the elements to be included in the QR code. All characters in $U$ and $T$ are assumed to be single-byte encodable using UTF-8 encoding.}
    \label{tab:lengths}
\end{table}

\section{Security Analysis}
\label{sec:security-analysis}

In this section we analyze our solution's security guarantees. We examine the two main building blocks of \approach{} and, for each of them, describe how our system defends against different (potential) attacks which follow from the threat model described in Section~\ref{sec:sys-n-attacker}.

\subsection{Security of the Image Comparison Algorithm}

It is challenging to mathematically describe all possible uncontrolled scenarios for smartphone videos. In the following, we explain why our algorithm will detect forgeries in realistic situations. For this analysis, we assume that two digital documents are input to our image comparison algorithm: a frame in the video of the document received by the user and a reference image of the (original) document. 

Let $d$ be an original document and let $\mathfrak{p}$ be a document containing a forgery $\alpha$. By the definition of forgery, the size of $\alpha$ is $|\alpha| > \tau$. Without loss of generality, let $|\alpha|< |\sigma|$   (the argument is similar otherwise). Let $d'$ be the (potentially distorted) picture of $\mathfrak{p}$ and assume that the output of Algorithm~\ref{alg:algo} does not contain $\alpha$. There are two cases when this can happen: \textbf{(a)} $\alpha$ was not detected in the first global difference calculation (line 8), and thus the algorithm never enters the loop (line 9) or \textbf{(b)} $\alpha$ is found initially, but in some round $k$ it is deleted from $\Delta$.

\paragraph{Case (a)} In this case, sufficiently many characteristic points of $d$ were detected in $d'$ and thus a homography was found, as otherwise the algorithm would have returned $\Delta = \{ (0,0),(n,m)\}$, which trivially contains $\alpha$. If $\alpha$ is not contained by any rectangle in the first estimation of $\Delta$ then the difference map of $d$ and $d'$ does not contain a connected set larger than $\tau$ in a vicinity of $\alpha$. This could arise in practice due to extreme lighting conditions. However this would likely also affect other parts of the document thus raising many other differences, suggesting to human users that the picture should be retaken. The effects of folding would also impact other areas of the document given that paper is non-stretchable, making it unlikely that a given forgery is invisible to the algorithm while keeping the differences in the rest of the document below the $\tau$ threshold in the difference map.

\paragraph{Case (b)} In this case, assume $\alpha$ is contained in some rectangle $r_i$ in $\Delta$ in the first round of Algorithm~\ref{alg:algo}. When recomputing a homography in the vicinity $\sigma$ of $r_i$, two cases are possible. First, a homography is found, corresponding to $d(\sigma)$. In this case, this homography will at most rotate $d'(\sigma)$ by an angle $\phi$ and adjust the perspective slightly. These operations will approximately preserve the area  $|\alpha|$. Rotation preserves the area of contours in a region while perspective adjustment could slightly increase or decrease the view of $\alpha$ in $d'$. However, local approximations will look more similar to the original (flat) subregion and thus in general $\alpha$ will look closer to a perfect scan (where $|\alpha| > \tau$). When recomputing the difference, $\alpha$ will thus be still present in $\Delta$ for the next round.
A second possibility is that no homography can be found that is good enough. In this case, $d(\sigma)$ will be compared against $d'(\sigma)$, which implies the original $r_i$ will be found again and passed along to the next round in $\Delta$.

This argument can be applied inductively to show that $\alpha$ will be contained by some $r_i$ in the fixed point. An exception would be a homography that would shrink the area of $\alpha$ below $\tau$. Given that $|\alpha| > \tau$ in a perfect scan of $\mathfrak{p}$, this means that the whole region $d'(\sigma)$ is transformed incorrectly, which implies other visible elements in $\sigma$ do not preserve their shape and are likely to cause noticeable (i.e. bigger than $\tau$) differences.

In practice, it is important to choose $\sigma$ and $\phi$ so that the family of equivalent scans does not alter the document semantics. For instance, choosing $\phi = \pi/2$ could accept semantics altering forgeries, such as an inverted chart, whereas $\phi = \pi/20$ is less likely to do so.\footnote{Unless perfect angles are crucial for the document in question, for instance in a building plan. In that case flatbed scans would be a better acquisition technique.}

\emph{From single frames to video} As discussed in Section~\ref{sec:requirements}, the verdict on a document video will be reached once $k$ consecutive individual frame analyses reach the same verdict. Following the security analysis of the image comparison algorithm for individual frames, the final video verdict will contain all forgeries detected consistently in $k$ consecutive frames. This lowers the likelihood of missing forgeries due to random lighting or focus anomalies, and otherwise preserves the security argument for individual frames. 

\subsection{Security of the Reference Document}
\label{sec:crypto-auth}

We next discuss the security guarantees that our system provides. We start by formalizing \approach{}'s main security goal. 

\begin{definition}[Document Origin Authentication]
\label{def:property}
For every issuer $I$ with private key $k_I$ and public key $K_I$, all byte-strings $U,\kappa,T,t$, and all $\tau,\delta,\phi,\sigma\in \mathbb{N}$, and all $d,d'\in \mathcal{D}$ such that:
\begin{itemize}
\item $d'$ and $d$ are $(\tau,\delta,\phi,\sigma)$-\emph{similar},
\item $d'$ contains the signature $Sign_{k_I}(h(U,h(d),\kappa,T,\tau,\delta,\phi,\sigma))$ in its QR code,
\item $\{d\}_{\kappa}$ is hosted at the URL $U$,
\item If $K_I$ has been revoked, then the timestamp $t$ corresponds to $S$ and to a time older than the revocation timestamp in the CRL.
\end{itemize}
it holds that either $d'$ is a picture of a document issued by $I$, or both the private key of the issuer $k_I$ \textbf{and} the server hosting $\{d\}_{\kappa}$ at $U$ have been compromised.
\end{definition}

We have produced a formal, computer-checked proof
that property \ref{def:property} holds for our design, with respect to our attacker model, using the Tamarin model-checking tool~\cite{tamarin13}. Tamarin is a symbolic model-checker for security protocols and we use it to model the actions and interactions between different system components, including the attacker compromising them and learning their associated keys.  

Following our attacker model, we model the physical channel between issuers and end users as an insecure channel, where tampering and spoofing of the document or the QR code can occur. We also model key $k_I$ compromise and revocation, which corresponds to the adversary gaining control of the issuer's long term secret. The compromise of the storage service means that attackers can upload and replace arbitrary documents on this service. Note that we do not model the compromise of the end user app; if this were compromised then the game is over as an attacker could falsely authenticate any document.

Tamarin produces a proof of this property automatically under one second on a commodity laptop. Intuitively this property holds because there is a kind of defense in depth where individual compromises are insufficient. For example, compromising the issuer's private key $k_I$ is not sufficient to forge a document, since if no corresponding digital version is stored on the storage service, the app will not be able to carry out the authentication. Moreover, compromising the storage service and only uploading a forged document without generating a valid QR with the key $k_I$ will be detected by the app when verifying the signature $S$. Finally, if we consider the compromise of the the timestamping service, which is equivalent to ignoring the timestamp coherence check in the model, our approach is secure unless the attacker additionally compromises both the issuers key $k_I$ and the storage service. This result validates our design in the presence of strong adversaries who can compromise some, but not all, of the system components.

We provide next an additional intuition on potential attack vectors and describe how they are prevented by our system design.


\paragraph*{Attacker forges QR and redirects to its own URL} This attack is detected by the system if there is no matching trusted public key $K_I$ for the legitimate issuer $I$ that can verify the signature $S$. This attack will only succeed if the attacker can obtain a digital signature for a trusted (i.e. not revoked) public key $K_I$ as well as compromising the storage service before $K_I$ is added to the Certificate Revocation List. 

\paragraph*{Attacker replaces reference document in storage service} Assume an attacker manages to compromise the storage service and replaces the (encrypted) reference document $Enc_{\kappa_d}(d)$, pointed to by $U$, by some other version $Enc_{\kappa_d'}(d')$. This attack will fail unless the attacker can generate a valid QR code matching the new $\kappa_d'$ and the new document hash $h(d')$, for which they would also need to compromised the issuer's key $k_I$.

\paragraph*{Other attacks} 

As explained previously, we assume during setup that public keys are distributed authentically. Concretely, consider the bank issuer example in the introduction. In this setting, users have a legitimate app installed on their phones with an authentic issuer's public key pinned. As a result, phishing attacks in the form of a malicious app are not possible because the app is delivered to users using trusted channels according to our system model.





\section{Evaluation}
\label{sec:evaluation}

In this section we evaluate of our forgery detection algorithm, which is at the core of our approach. To do so, we implemented Algorithm~\ref{alg:algo} in Python using the OpenCV library.  We implement traditional preprocessing techniques to remove shadows and improve content contrast, such as adaptive thresholding.

\begin{figure*}[tb]
    \centering
    \includegraphics[width=\textwidth]{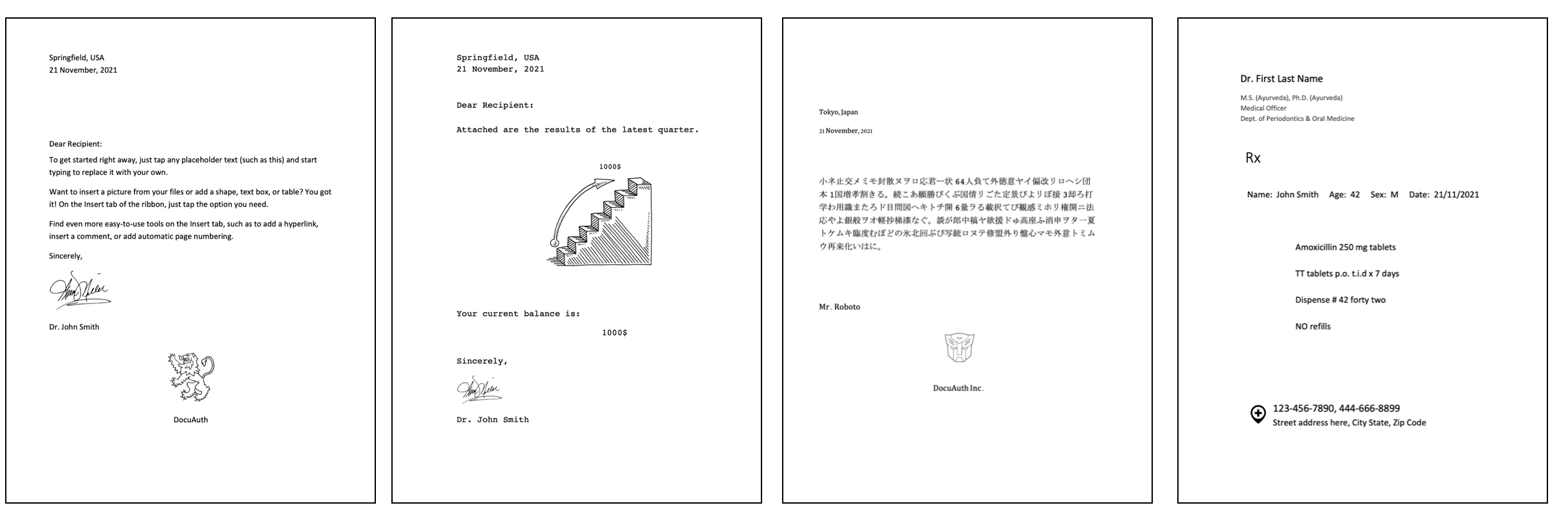}
    \caption{Four document types: a business letter, a quarterly result letter, a business letter in Japanese, and a medical prescription.}
    \label{fig:samples}
    
\end{figure*}

\begin{figure*}[tb]
    \centering
    \includegraphics[width=0.6\textwidth]{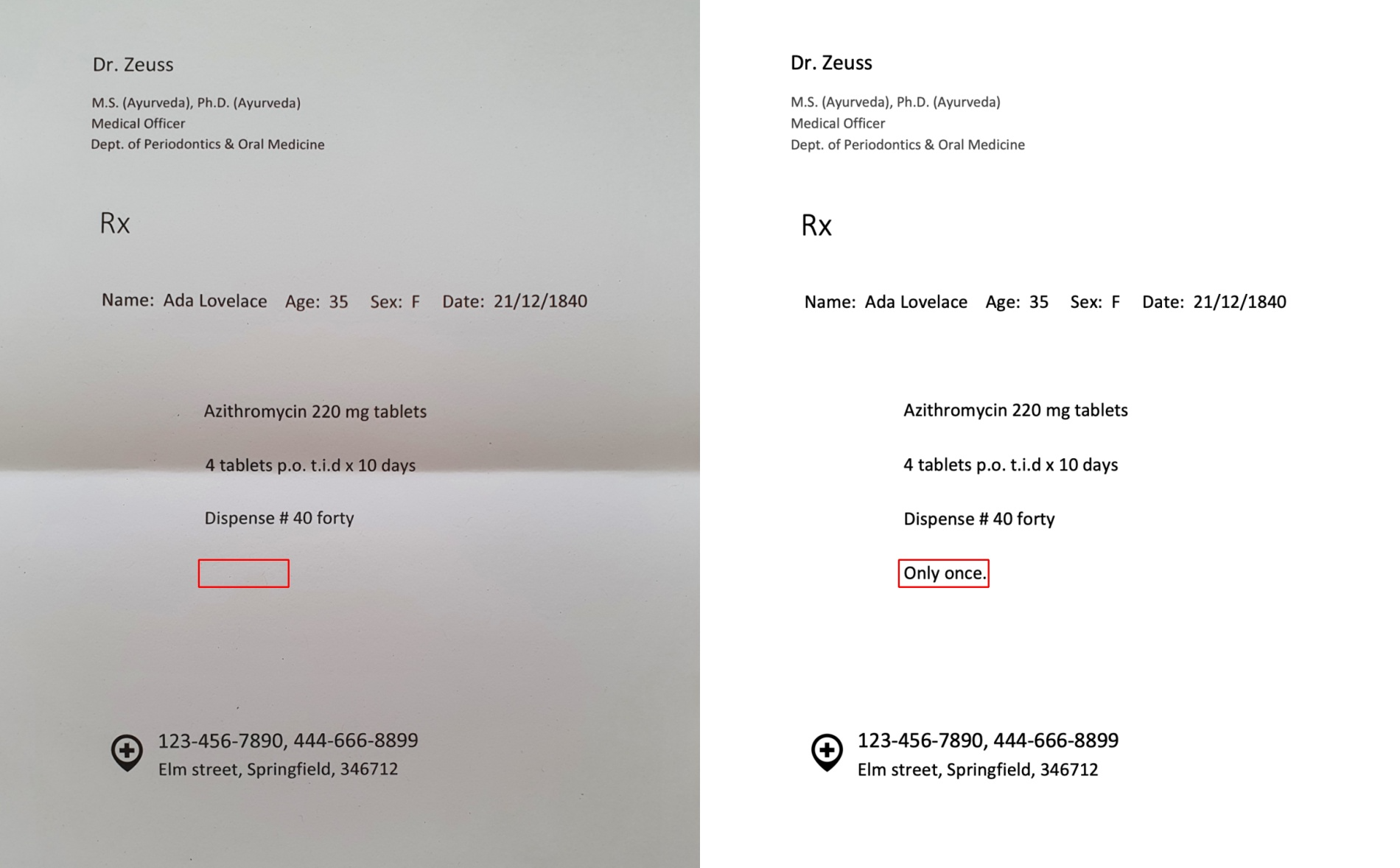}
    \caption{Example analysis outcome of the algorithm on document containing forgery.}
    \label{fig:analysisexample}
    
\end{figure*}


\paragraph{Dataset} We evaluate our prototype's performance on a set of 128 videos taken from 32 printed documents with 4 different smartphones, including 2 Android (Xiaomi redmi note 8 and 11) and 2 iOS devices (iPhone 6s and 12pro). Each video has a duration of approximately 5 seconds. Videos were taken under different lighting conditions and printed documents were either unfolded or previously folded in halves or thirds. Figure \ref{fig:samples} illustrates examples of 4 templates used for the document generation. For each of these document classes, there are 4 legitimate documents with different contents, dates, and signatures. For each of the 16 legitimate document, we produce a forgery, ranging from very subtle (e.g.~a minus sign before a number or altering one digit in a date) to relatively noticeable (e.g.~removing a sentence or changing the name of the signing individual).

\paragraph{Parameters} Before analysing the videos of the printed documents, for each document class we pick a value for $\delta$ and $\tau$ such that the digital versions of the 4 forgeries in this class are correctly detected by the algorithm. This yields slightly different but similar values for all classes, which is explained by the fact that they have different font types, sizes, and character sets. The values of $\tau$ range from 10 to 20 pixels, which is about half the size of a comma in the reference document resolution (1700px $\times$ 2200px). The value of $\delta$ is the same for all documents at $0.35$.

We set $\sigma$ to be the same for all documents at 80px $\times$ 80px and for efficiency's sake we only perform two refinement rounds instead of computing a fixed point. Empirically we observe that additional rounds do not significantly improve the analysis for our dataset. We set $\phi = 30^\circ$.

\paragraph{Results} Our analysis annotates a total of 100 regions as potential forgeries. These correspond exactly to all forgeries in the documents used for the evaluation. That is, for each rectangle containing a forgery in the ground truth, our method finds a rectangle that non-trivially intersects it. This intersection is on average 99\% of the area of the ground truth rectangle. 
The average size of the rectangles in the ground truth is 0.25\% of the total document area, whereas the area of the estimated rectangle by the algorithm is on average 0.94\% (i.e. less than 1\% ).
This $\sim$ 0.7\% difference in area sizes is because,  to improve usability, we augment difference rectangles to merge those rectangles that are very close to each other into a single rectangle.

False alarms could arise when there are misaligments with respect to the original documents that could not be resolved by our iterative technique, for instance because no good homography was found for that document region. Although sporadic false alarms occur in some frames of our videos, the convergence analysis for $k=3$ prevents them from occurring in the final reported result.

Figure \ref{fig:analysisexample} shows a paper document and an original document together with the automatically added annotations. The running time of the individual frame analysis ranged from 0.5 to 1 second, depending on the number of initially estimated differences in the algorithm's first round. For our evaluation, we perform an offline analysis at a rate of 4 frames per second. The rationale behind  this sampling rate is that successive images should be sufficiently distinct  but the analysis should converge as fast as possible. The convergence analysis reaches a decision on average after 5.13 frames (standard deviation 3.66), which corresponds to 1.28 seconds of video. 


The results of this evaluation show that our algorithm is accurate enough to assist human users in identifying forgeries. Namely, it detects and annotates all forgeries, the false alarms are very rare and the annotations are meaningful to end users who can then manually check these rare cases. No false alarms arose in our evaluation, however sporadic false alarms appeared on individual frames, which were  corrected by the frame coherence analysis. 


Our evaluation has some limitations. First, the number of devices and pictures taken, although providing evidence for the generalizability of the approach, can still be enlarged to cover more scenarios. Videos in our evaluation were realistic in that we experimented with various lighting conditions and documents were folded. However, in practice, users may produce videos under even more challenging conditions, which  would potentially produce more false alarms (due to shadows, lack of focus, stains etc.). Thus, a comprehensive user study where participants are asked to take their own videos on their devices should be designed and carried out; this is left for future work.

\section{Related Work}
\label{sec:relatedwork}

The idea of extending paper documents with characteristics that can facilitate digital cryptographic authentication has been discussed by Eldefrawy et al.~\cite{eldefrawy2012hardcopy}, Wang et al.~\cite{wang2015cryptopaper}, and Li et al.~\cite{li2015authpaper}. In contrast to these works, our solution considers both off-line and on-line techniques, provides an algorithm for automatic image comparison, and gives a detailed analysis of the security guarantees of the approach. 

In the following we describe and compare against other approaches that are related to the building blocks used in \approach{}.

\paragraph{OCR-based authentication}

An alternative is to relay directly on optical character recognition (OCR) and compare character streams.  Unfortunately, OCR also has accuracy issues \cite{accuracyOCR} (stemming from problems similar to those addressed in this work, such as lighting, foldings and overall picture quality). Moreover, OCR suffers from several drawbacks compared to our approach. OCR assumes known  character sets upon which it is trained, whereas our forgery detection method is more general. For instance we can detect differences in new alphabets or character fonts. Moreover, we can also account for pictures and characters' positions, both of which are relevant for the document's semantics. 

 Ambadiyil et al.~\cite{qrwithocr} present an approach that can be verified using OCR technology by hashing critical parts of a document and  other random segments of text. Our approach is however more general as it can detect forgeries throughout the document and critically on its non-textual parts.

\paragraph{Machine-learning based approaches}
The research most closely related to our image comparison technique is~\cite{andreeva2020comparison}.  This work compares scanned administrative documents and highlights visual differences to detect forgeries. Two fundamental differences with our approach are that their approach is tailored for textual documents, whereas ours is more general, and that they rely on high quality flatbed scans.  Works such as Kim et al.~\cite{kim2018detection} use machine learning to characterize differences between documents with a focus on purely textual (digital) documents and version managing.

It is  challenging  to attain generalizability of a deep neural network trained on a dataset containing authentic documents and forgeries to achieve a reliable image comparison algorithm, given that one may not consider all possible interesting forgeries or types of authentic documents. Moreover, one would have to take into account adversarial attacks against such solutions, which is still a topic of ongoing research. In our setting, we can mathematically characterize attacks and thus provide precise accuracy guarantees. However, situations may arise that are difficult to handle by our algorithm, such as attacks close to the size of $\delta$ or the intensity threshold $\tau$, together with particular lighting conditions. Our evaluation shows that our algorithm supports choices of parameters that make such attacks difficult, given that the size (smaller than a comma) and intensity (similar to shadows) would make them unlikely to impact a document's semantics.

\paragraph{Other forgery detection approaches}
 Van Beusekom et al.~\cite{van2008document} propose a technique that creates signatures out of known authentic documents. This technique can detect forgeries where attackers scan and reprint original documents, since in this process distortions of known authentic subregions of the document tend to occur. This approach however would not detect sophisticated forgeries that, for instance, simply add a comma to an authentic document.

Similarly, the work of Picard et al.~\cite{picard2004digital} on copy detection patterns raises the bar against forgery under the assumption that copying/reprinting variants of an original document also degrades their quality. Although this technique can detect some forgeries, it cannot pinpoint the specific regions that have been tampered with and its security guarantees are unclear~\cite{9648384}.

\paragraph{QR extensions for authenticating paper documents}
The Federal Office for Information Security in Germany has proposed colored QR codes with the goal of extending their storage capacity and has used them in physical document authentication~\cite{jabcode}. Similarly, Yang et. al.~\cite{yang2018robust} propose high capacity QR codes. In theory these approaches can encode several KBs of data in a single QR. However, our experiments with open source implementations only allowed us to reliably encode at most 2KB. If technology in smartphones (i.e. camera resolution) improves in the future, our technique could be adapted to work fully offline.

\paragraph{Document dewarping}
Document dewarping research~\cite{dewarping1,dewarping2} is related but orthogonal because it does not assume knowledge of ground truth (the authentic image). Often the goal of document dewarping is to improve OCR. In contrast, our goal is more general since forgeries could also happen in graphics like charts, signatures, and other images that cannot be parsed using OCR.

\paragraph{Alternative image similarity functions} It is possible to instantiate our image comparison approach with alternative image similarity functions, such as the structural similarity index (SSIM, \cite{wang2004image}). SSIM was designed to measure visible differences between various versions of an image to judge, for instance, how much would noise generated by a compression process affect the quality of a picture from a human user's point of view. We have preliminary experiments with SSIM in the context of document authentication, and although it offers advantages for reducing false positives, it poses challenges for forgeries that overlap content present in the original image. Moreover, we found that the absolute difference map provides a more intuitive explanation of such forgeries from a security point of view in the context of document forgery.

\paragraph{Image change detection algorithms} Algorithms  have also been proposed for image change detection \cite{radke2005image}. Such algorithms are useful to detect changes in the same scene, for example in the context of video surveillance (where new persons or objects enter the scene). There is however a fundamental difference with respect to our problem setting: although in principle the authentic document is a ``scene'' , it constitutes only one frame and (substantial) differences in the to-be-verified video will be typically introduced by paper foldings, lighting conditions, and adversarial changes. Thus, for both security and accuracy reasons, we found it preferrable to develop an approach exploiting the characteristics of our problem setting (natural non-adversarial foldings and simple shadow pre-processing). We leave a detailed comparison with image change detection algorithms as interesting future work.

\paragraph{Non-repudation and revocation} It is well known that key revocation is problematic for non-repudation  in the context of digital signatures~\cite{nonrepudiation2}. Resolving this tension requires additional mechanisms such as secure timestamping \cite{nonrepudiation1} or forward secure signature schemes~\cite{bellare1999forward, nonrepudiation2}. In \approach{}, we have used secure timestamps to verify the chronological order of events, thus avoiding the necessity of repudiating all documents signed prior to a key being revoked due to compromise. Commercial services such as \cite{surety} provide similar time-stamping guarantees. In our setting, a malicious storage administrator can delete the reference image corresponding to a given issued document, thus de-facto repudiating it. A decentralized storage solution would mitigate this threat, but its formal treatment is left for future work.

In sum, to the best of our knowledge, we are the first to propose a comprehensive approach to automatically verify the authenticity of rich paper documents with the help of smartphones by using image comparison and cryptography.

\section{Conclusions}
\label{sec:conclusions}

 We have presented \approach{}, a novel approach that uses cryptographic and image comparison techniques  to authenticate rich paper documents containing both text and graphics. Our solution provides rigorous security guarantees which are verified formally. A preliminary evaluation of \approach{} shows that it significantly raises the bar against forgery in challenging  scenarios. In the future, we plan to perform a large-scale user study to further suppport our claims and gain more insights regarding its usability. 

\section*{Acknowledgements}  We thank the Werner Siemens-Stiftung for their generous support of this project, under the Centre for Cyber Trust.

\bibliographystyle{ACM-Reference-Format}
\bibliography{biblio}

\end{document}